\documentclass[conference]{IEEEtran}
\usepackage{amsmath}
\usepackage[boxed,algosection,linesnumbered]{algorithm2e}
\usepackage{epsfig}

\ifCLASSINFOpdf
\else
\fi
\begin{document}
\title{Self-Adaptive Consolidation of Virtual Machines For Energy-Efficiency in the Cloud}

\author{\IEEEauthorblockN{Guozhong Li, Yaqiu Jiang,Wutong Yang, Chaojie Huang}
\IEEEauthorblockA{School of Information and\\ Software Engineering\\
University of Electronic Science \\and Technology of China\\
}

\and
\IEEEauthorblockN{Wenhong Tian}
\IEEEauthorblockA{School of Information and\\ Software Engineering\\
University of Electronic Science \\and Technology of China\\
Email: tian\_wenhong@uestc.edu.cn}}

%Telephone: (800) 555--1212, Fax: (888) 555--1212}


\maketitle

\begin{abstract}
In virtualized data centers, consolidation of Virtual Machines (VMs) on minimizing the number of total physical machines (PMs) has been recognized as a very efficient approach. This paper considers the energy-efficient consolidation of VMs in a Cloud Data center. Concentrating on CPU-intensive applications, the objective is to schedule all requests non-preemptively, subjecting to constraints of PM capacities and running time interval spans, ~such that the total energy consumption of all PMs is minimized (called MinTE for abbreviation). The MinTE problem is NP-complete in general. We propose a self-adaptive approached called SAVE. The approach makes decisions of the assignment and migration of VMs by probabilistic processes and is based exclusively on local information, therefore it is very simple to implement.
Both simulation and real environment test show that our proposed method SAVE can reduce energy consumption about 30$\%$ against VMWare DRS and 10-20$\%$ against EcoCloud on average.
\end{abstract}


% \fi
\IEEEpeerreviewmaketitle

\section{Introduction}
Cloud computing has evolved from various recent advancements in virtualization, Grid computing, Web computing, utility computing and other related technologies. %Cloud computing provides both platforms and applications on demand through the Internet or intranet. Cloud computing allows the sharing, allocation and aggregation of software, computational and storage network resources on demand. Some of the key benefits of Cloud computing include the hiding and abstraction of complexity, virtualized resources and efficient use of distributed resources. Some examples of emerging Cloud computing platforms are Google App Engine [23], IBM blue Cloud [24], Amazon EC2 [21], and Microsoft Azure [25]. Cloud computing is still considered in its infancy as there are many challenging issues to be resolved. Youseff et al. [20] establish a detailed ontology of dissecting Cloud into five main layers from top to down: Cloud application (SaaS), Cloud software environment (PaaS), Cloud software infrastructure (IaaS), software kernel and hardware (HaaS), and illustrates their interrelations as well as their inter-dependency on preceding technologies.\\
It offers three level of services, namely Infrastructure as a Service (IaaS), Platform as a Service (PaaS) and Software as a Service (SaaS).
In this paper, we concentrate on CPU-intensive computing at Infrastructure as a Service (IaaS) level in Cloud Data centers. Cloud computing providers (such as Amazon) offer virtual machines (VMs) with specified computing units.
%\section{The Scheduling of Virtual Machine requests in a Cloud Data center}
The resources in this paper include:
\begin{enumerate}
\item
Physical Machines (PMs): physical computing devices which can host multiple virtual machines; each PM can be a composition of CPU, memory, hard drives, network interface cards (NICs), and etc..
%\item
%physical clusters (PCs): consist of a number of PMs, necessary network and storage infrastructure.\newline
\item
Virtual Machine (VMs): virtual computing platforms on PMs using virtualization software; each VM has a number of virtual CPUs, memory, storage, NICs, and related components.
%\item
%virtual cluster (VCs): consist of a number of VMs and necessary network and storage infrastructure.\newline
\end{enumerate}

%\subsection{Scheduling Process in a Cloud Data center}
The architecture and process of VM scheduler are provided in Fig.~1., referring to Amazon EC2 \cite{Amazon}. As noted in the diagram, the major processes of resource scheduling are:
\begin{enumerate}
\item
User requesting: the user initiates a reservation through the Internet (such as Cloud service provider's Web portal);
\item
Scheduling management: Scheduler Center makes decisions based on the user's identity (such as geographic location, etc.) and the operational characteristics of the request (quantity and quality requirements).
The request is submitted to a data center, then the data center management program submits it to the Scheduler Center, finally the Scheduler Center allocates the request based on scheduling algorithms;
\item
Feedback: Scheduling algorithms provide available resources to the user;
\item
Executing scheduling: Scheduling results (such as deploying steps) are sent to next stage;
\item
Updating and optimization: The scheduler updates resource information, optimizes resources in the data center according to the optimizing objective functions.

\end{enumerate}

\begin{figure} [htp!]
\begin{center}
\hfill
{\includegraphics [width=0.45\textwidth,angle=0] {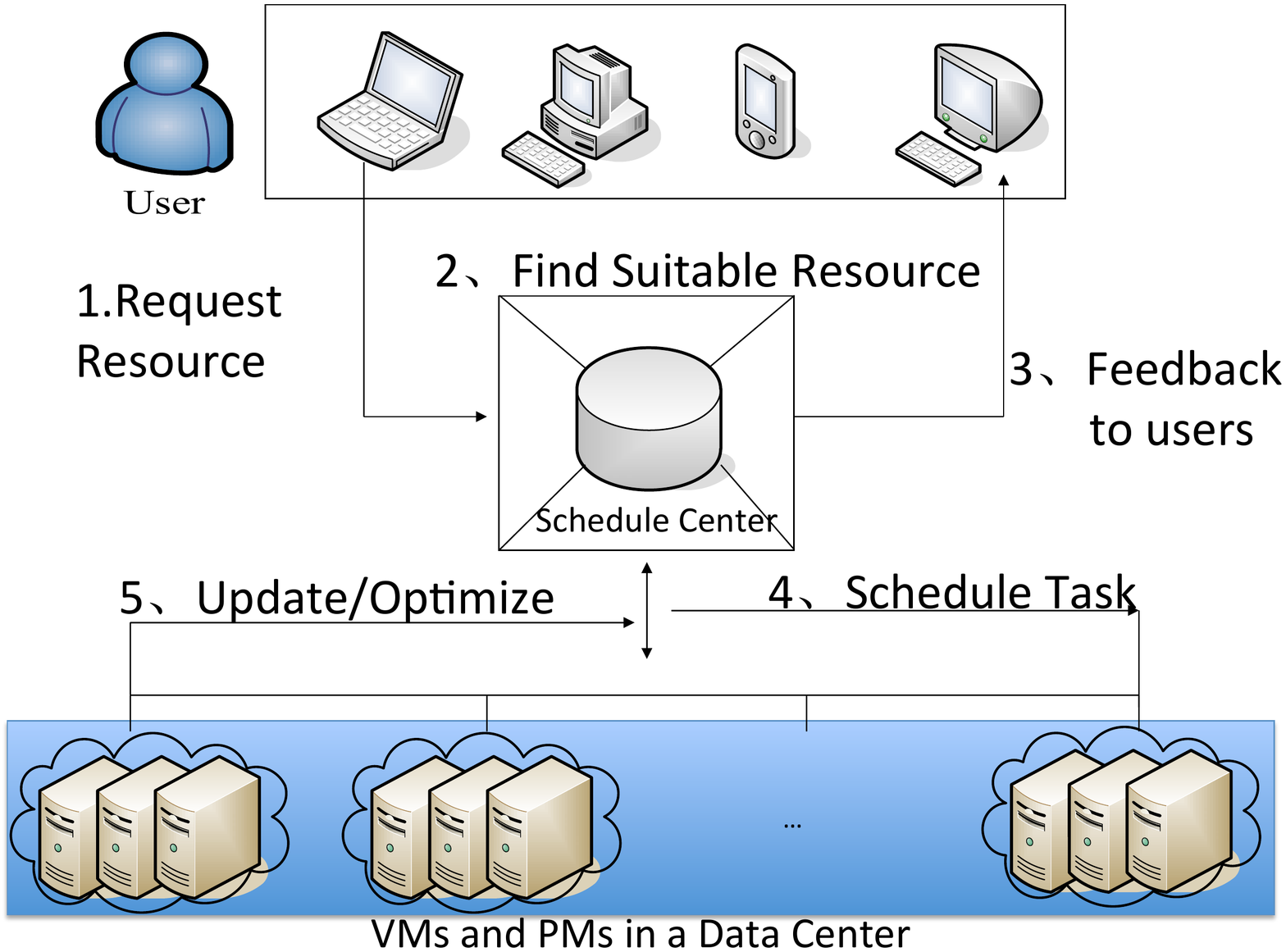}}
\hspace*{\fill}
\caption{Referred architecture of VM consolidation in a Cloud data center}
\end{center}
\end{figure}

In Cloud service, customers are billed in a way proportional to the total amount of computing time as well as energy of the computing resources. The scheduler executes periodically for a fixed period of time, for instance, every 15 minutes or so depending on workloads in realistic scenarios. From providers' point of view, the total energy cost of computing resources is closely related to the total number of PMs used and total powered-on time of all PMs.
In practice, idle server consumes between 60 and 70 percent of the total power and energy consumed when it is fully utilized. VM consolidation aims to alleviate this problem. It tries to allocate a set of VMs using the minimum number of PMs and also migrate VMs when the hosting PM has CPU utilization below the predefined lower bound or beyond the upper bound. However, the optimal allocation of VMs to PMs is a NP-hard problem [2][15]. In this paper we propose heuristic algorithms for both VM allocation and migration. When a Cloud data center becomes very large, traditional centralized allocation and migration face great challenge. The centralized method can be time consuming and affecting the efficiency, and existing methods require the simultaneous migration of many VMs (such as in [2]) to reduce energy consumption, this can cause massive overhead, service degradation or network hibernating. To alleviate these issues, we propose self-adaptive allocation and migration algorithms for VM consolidation. They combine the advantages of centralized and decentralized approaches by setting allocation probability and migration probability function in each PM, and the scheduler collects the information from each PM (or a set of small number of PMs) to allocate and migrate VMs. Therefore they reduce the overhead, network vibrating, and total energy consumption in a data center.

As Cloud data centers consume very large amount of energy and the energy cost (electricity price) is increasing regularly. So they like to minimize total power-on time of all PMs used to save energy costs. In practice, some simple algorithms (such as Round Robin and First-Fit) are used by EC2 \cite{Amazon} and VMWare \cite{VMWare}. Currently there is still lack of energy-efficient scheduling algorithms.

The problem of VM consolidation can be stated generally as follows. There are $n$ deterministic VM requests submitted to the scheduler to be scheduled on multiple PMs with bounded capacities. Each VM request (job) is associated with a start-time, an end-time, and a capacity demand. The objective is to schedule all requests non-preemptively, subjecting to constraints of PM capacities and running time interval spans, such that the total energy consumption of all PMs is minimized (called MinTE for abbreviation).

In this study, we assume that the total CPU capacity of a PM, $g$, is measured in abstract units such as EC2 Compute Unit (ECU)\footnote{The EC2 Compute Unit (ECU) provides the relative measure of the integer processing power of an Amazon EC2 instance and provides the equivalent CPU capacity of a 1.0-1.2 GHz 2007 Opteron or 2007 Xeon processor}.
Tian et al.\cite{Tian2013b} propose a 3-approximation algorithm called MFFDE for general offline parallel machine scheduling with unit demand, the MFFDE applies FFD with earliest start-time first.
\textit{The jobs and VM requests are used interchangeably in this paper}.\\ The major \textbf{contributions} of this paper include:
\begin{enumerate}
\item
Proposing a self-adaptive VM allocation and migration algorithms for energy-efficient scheduling.
\item
The proposed method SAVE can efficiently reduce total energy consumption compared against VMWare DRS (DPM) [12].
\item
Conducting intensive simulation and real environment tests.
\end{enumerate}
The rest of the paper is organized as follows. Formal problem statement is provided in Section \ref{problem formulation}. Section \ref{metrics} considers how our results are applied to the energy efficiency of VM requests. Performance evaluation is conducted in Section \ref{performance}. Related work is discussed in Section \ref{related work}. Finally we conclude in section \ref{conclusion}.



\section{Problem Formulation}
\label{problem formulation}
\subsection{Preliminaries}
%\begin{definition} \label{Definition 3}
For energy-efficient scheduling, the objective is to meet all VM requests with the minimum total energy consumption based on the following assumptions and definitions.\newline
\begin{enumerate}
\item
All data are deterministic unless otherwise specified, the time is discrete in slotted window format. We partition the total time period [0, T] into slots of equal length ($l_0$) in discrete time, thus the total number of slots is $k$=$T/l_0$ (always making it a positive integer). The start-time of the system is set as $s_0$=0. Then the interval of a request $i$ can be represented in slot format as a tuple with the following parameters: [StartTime, EndTime, RequestedCapacity]=$[s_i, e_i, d_i]$. With both start-time $s_i$ and end-time $e_i$ are non-negative integers.
\item
For all jobs, there are no precedence constraints other than those implied by the start-time and end-time. Preemption is not considered.
\item
The performance of SAVE is assessed through the following metrics:\\
a). Resource utilization: To foster consolidation and save energy, a server should be either highly exploited or in a sleep mode. Analysis of CPU utilization aims at checking if this objective is
fulfilled. \\

b). Number of active servers: VMs should be clustered into
as few servers as possible. For example, if the overall load of the data center is about 50 percent of the total available capacity of PMs, the number of active PMs should be close to 50 percent of the overall number of PMs.\\

c). Energy consumption: The power and energy consumed by the whole data center in different load conditions are computed. \\

d). Number of VM migrations: Though migrations are essential for VM consolidation and energy reduction, it is important to limit their frequency to avoid massive overhead and reduce service degradation or downtime duration.

\end{enumerate}

\subsection{Allocation and Migration Functions}
The allocation procedure is performed when a user asks the data center for a new request. The request is considered as a VM, the data center scheduler must allocate the VM to one of the servers for execution.

Each server takes its decision whether or not to accept the request, consolidating the workload on as few servers as possible. The request should be rejected if the server is over-utilized or under-utilized on CPU. In the case of over-utilization, it is to avoid overload situation while in the case of under-utilization the objective is to put the server in a sleep mode and save energy, so the under-utilized server should refuse new VMs and migrate those that are currently running. Consequently, a server with highest utilization but not over-utilized should accept new VMs to foster consolidation.

To maximize the utilization of PMs, we define the allocation function for each PM as

\begin{equation}
f_a(x)=\frac{1-e^{-x}}{1-e^{-1}}, x\in [0,T_a]
\end{equation}
where $x$ is the current CPU utilization, $T_a$ is the upper threshold for CPU, and $f_a(x)$ is a monotonically increasing function with regard to to $x$.~If $x>T_a$, the corresponding PM will not accept any request. It can be seen that the higher of CPU utilization, the higher the acceptance probability (AP) will be.

\begin{figure}[h] %AnAccessNetwork.eps
\hfill\includegraphics [width=0.45\textwidth, angle=-0]{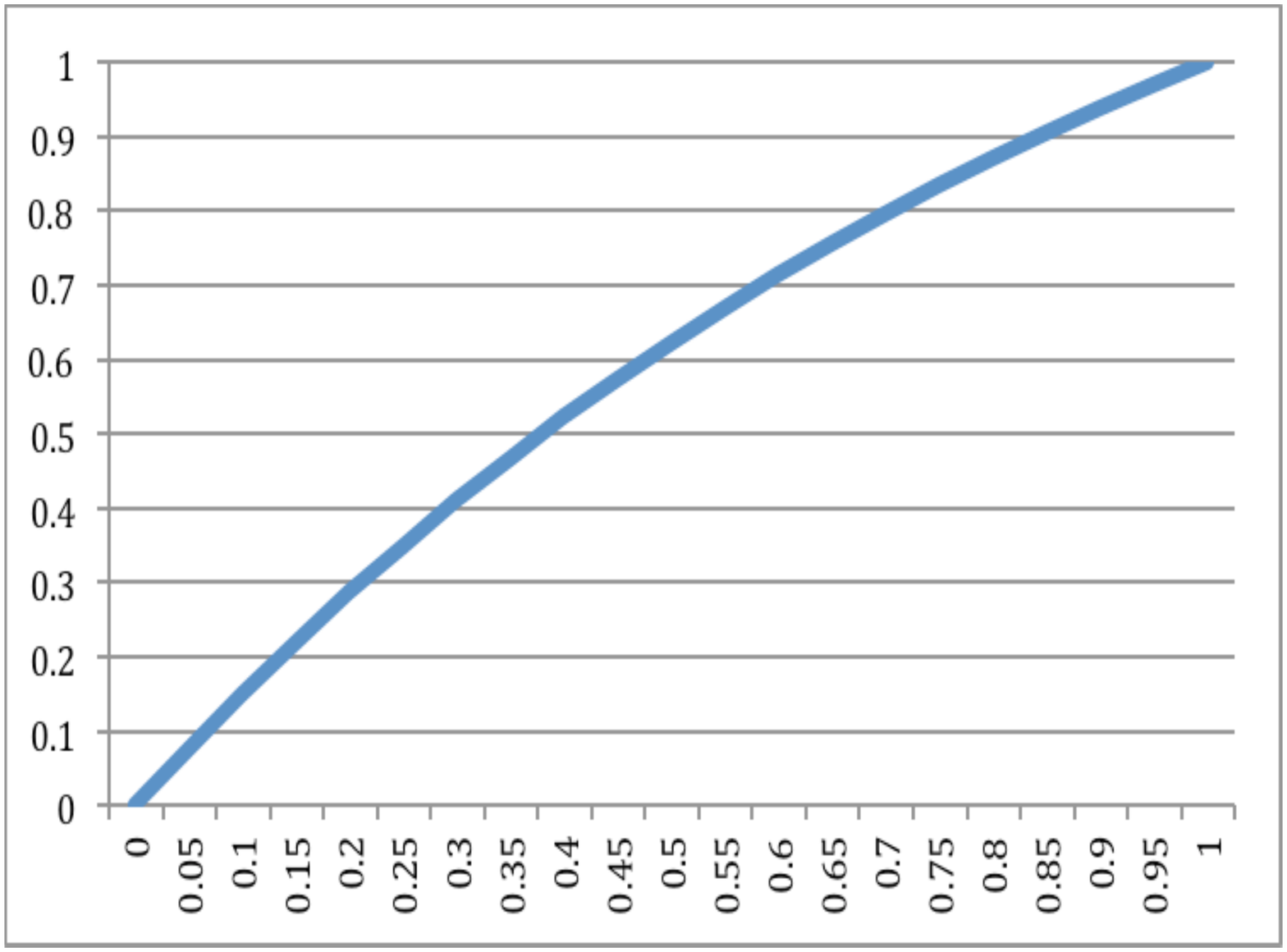}\hspace*{\fill}
\caption{The allocation function} \label{f5.1}
\end{figure}

\begin{algorithm}
\caption{SAVE-Allocation algorithm}
\label{algorithm_save}
\SetKwInOut{Input}{input}
\SetKwInOut{Output}{output}
\Input{A VM request, available PMs in the data center}
\Output{Scheduled job and hosting PM ID}
The scheduler collects acceptance probabilities (APs) from all PMs (or a set of PMs),
which compute APs following the Equ. (1)\;
Chooses the PM with highest AP to allocate current request \;
Updates the CPU utilization and available capacity of the PM\;
\end{algorithm}
The SAVE-Allocation algorithm has computational complexity of $O(N)$ where $N$ is the number of PMs considered.\\

The migration procedure is defined as follows: each server monitors its CPU utilization and checks if it is between two specified thresholds, the lower threshold $T_l$ and the upper threshold
$T_h$. When this condition is violated, the server evaluates the corresponding probability function, $f_m(x)$, and migrate
performs a modified Beta distribution whose success probability is set to the value of the function. We propose the migration function as
\begin{equation}
f_m(x)=\frac{-f(x,\alpha,\beta)}{3}+1, x\in (0,T_l]~or~x\in [T_h,1]
\end{equation}
where $T_l$ and $T_h$ is the lower bound and upper bound for CPU utilization respectively, and $f(x,\alpha,\beta)$ is Beta distribution defined as
\begin{equation}
f(x,\alpha,\beta)=\frac{x^{\alpha-1}(1-x)^{\beta-1}}{\int_{0}^{1} u^{\alpha-1}(1-x)^{\beta-1} du}
\end{equation}

\begin{algorithm}
\caption{SAVE-Migration algorithm}
\label{algorithm_save}
\SetKwInOut{Input}{input}
\SetKwInOut{Output}{output}
\Input{A number of $M$ PMs with high migration probabilities (MPs)}
\Output{the scheduled job, source and destination PM IDs}
The scheduler collects migration probabilities (MPs) from all PMs (or a set of PMs)\;
which compute MPs following the Equ. (2)\;
Chooses the PM with highest MP to migrate, the selected VM(s)
meet one of the following criterions: 1). the migration will empty the PM so that
the PM can be put into sleep mode to save energy \;
or 2). the migration makes the CPU utilization of the PM just below the upper bound
$T_h$ and only one VM is migrated \;
The selected VM(s) will be allocated according to SAVE-Allocation algorithm\;

\end{algorithm}

\begin{figure}[h] %AnAccessNetwork.eps
\hfill\includegraphics [width=0.45\textwidth, angle=-0]{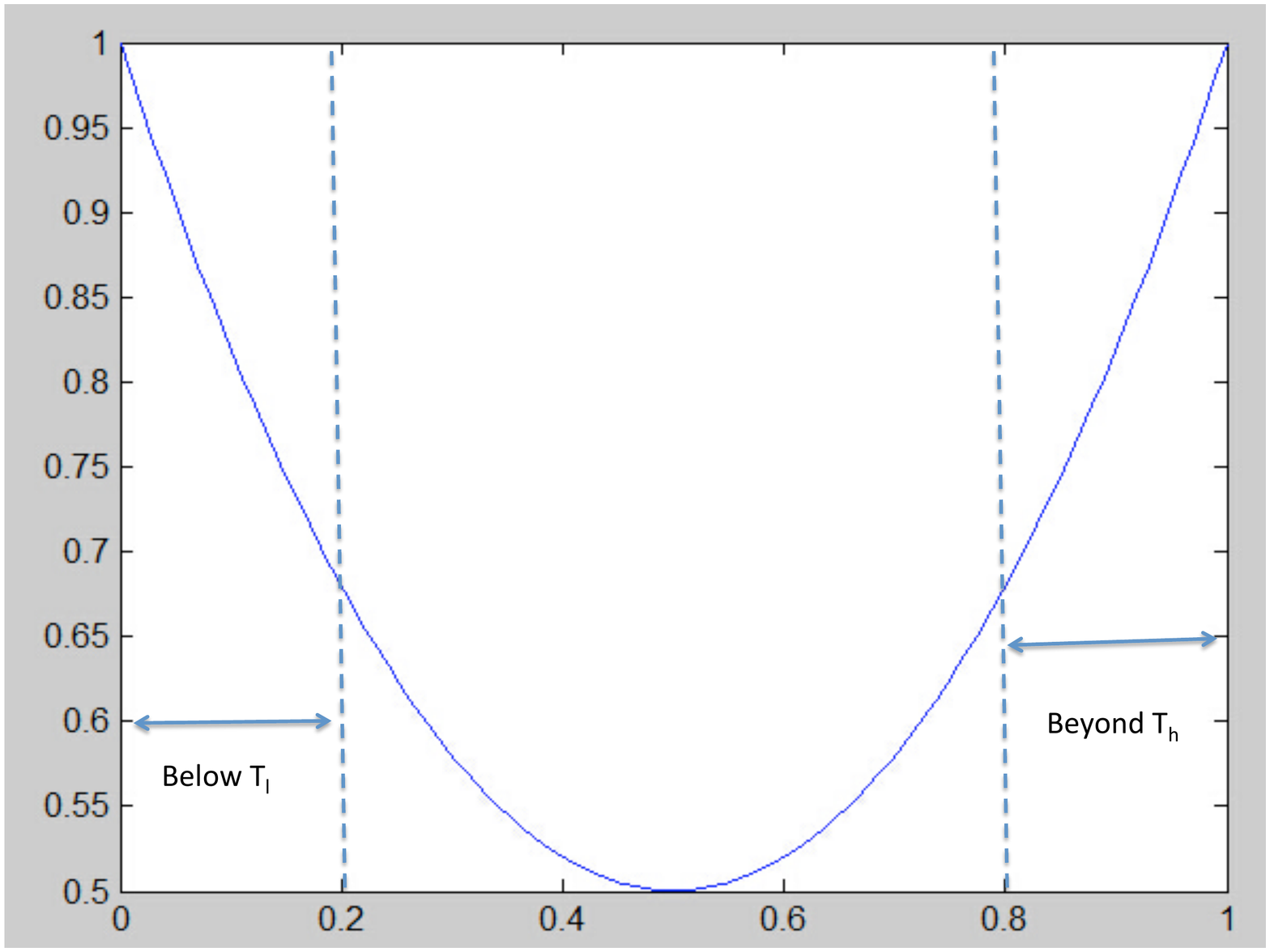}\hspace*{\fill}
\caption{The migration function} \label{f5.1}
\end{figure}
The migration function is shown in Fig. 3 where $\alpha$=2 and $\beta$=2. It is defined so as to trigger the migration of VMs when the utilization is below the threshold $T_l$ or above the threshold $T_h$, respectively. The shape of the functions can be modulated by tuning the parameters $\alpha$ and $\beta$, which can therefore be used to foster or hinder migrations.
VMs are periodically migrated to PMs which still can host by following the SAVE\_Allocation algorithm. \\
The SAVE\_Migration algorithm has computational complexity of $O(MK)$ where $M$ is the number of PMs
and $K$ is the total number of VMs considered for migrations, respectively.\\

\textbf {Observation 1. Our migration function ensures a gradual and continuous migration process by definition function (2), while most other techniques proposed for VM migration require the simultaneous migration of many VMs.}\\

This migration function is such proposed that gradual migration process can be controlled by choosing suitable number of PMs and the migration probability function. The aim is to avoid massive overhead, service degradation and network hibernating caused by simultaneous migration of many VMs.

\section{Metrics for energy-efficiency scheduling}
\label{metrics}
(1). The power consumption model of a server.\newline
There are many research works in the literature indicating that the overall system load is typically proportional to CPU utilization (see Beloglazov et al. [2], Matthew et al. \cite{Mathew2012}). This is especially true for CPU-intensive computing where CPU utilization dominates. The following linear power model of a server is widely used in literature (see for example [2][14] and references therein).
\begin{align}
P(U)&=kP_{max}+(1-k)P_{max} U\nonumber\\
&=P_{min}+(P_{max}-P_{min}) U
\end{align}
where $P_{max}$ is the maximum power consumed when the server is fully utilized, $P_{min}$ is the power consumption when the server is idle; $k$ is the fraction of power consumed by the idle server (studies show that on average it is about 0.7); and $U$ is the CPU utilization.
%This paper focuses on CPU power consumption, which accounts for main part of energy comparing to other resources such as %memory, disk storage and network devices.
In a real environment, the utilization of the CPU may change over time due to the workload variability. Thus, the CPU utilization is a function of time and is represented as $U_i(t)$. Therefore, the total energy consumption ($E_i$) by a PM can be defined as an integral of the power consumption function during [$t_0$,~$t_1$]:
\begin{equation}
E_i=\int_{t_0}^{t_1} P(U_i(t))dt
\end{equation}
When the average utilization is adopted, we have $U_i(t)=U_i$, then
\begin{align}
E_i&= P(U_i) (t_1-t_0)=P(U_i)T_i \nonumber \\
&=P_{min}T_i+(P_{max}-P_{min})U_iT_i
\end{align}
where $T_i$ is the power-on time of machine $PM_i$, the first term $P_{min}T_i$, is the energy consumed by power-on time of $PM_i$, denoted as $E_{i_{on}}$ (=$P_{min}T_i$); the second term, $(P_{max}-P_{min})U_iT_i$ is the energy increase by hosting VMs on it.
Assuming that a $VM_j$ increases the total utilization of $PM_i$ from $U$ to $U'$ and set $U'-U=u_{ij}$, and $VM_j$ works in full utilization in the worst case. Defining $E_{ij}$ as the energy increase after running $VM_j$ on $PM_i$ from time $t_0$ to $t_1$, we obtain that:

\begin{eqnarray}
E_{ij} &=& (P_{min}+(P_{max}-P_{min})U'- \nonumber \\
& &(P_{min}+(P_{max}-P_{min})U))(t_1-t_0) \nonumber\\
&=&(P_{max}-P_{min})(U'-U)(t_1-t_0) \nonumber\\
&=&(P_{max}-P_{min}) u_{ij} (t_1-t_0) %~~~~~~~~~~~~~~~~~~~~~~~~ ~~~~~~~~~~~~~~~~~~~~~~~~~~~(11)\\
\end{eqnarray}
For VM requests, we can further obtain that the total energy consumption of $PM_i$, it is the sum of its idle energy consumption ($E_{i_{on}}$) and the total energy increase by hosting all VMs allocated to it.
%The total power consumption of the $PM_i$ can be expressed as the sum of energy consumption when it is powered-on and the %energy consumed by all VMs allocated on it, formally, the energy consumption of the $PM_i$ is:\\
\begin{align}
E_i &= E_{i_{on}} + \sum_{j=1}^{k} E_{ij} \nonumber\\
&= P_{min} T_i + (P_{max}-P_{min}) \sum_{j=1}^k u_{ij} t_{ij} \label{eqn_E_i}
\end{align}
where $u_{ij}$ is the utilization increase of $PM_i$ with the allocation of $VM_j$, and $t_{ij}$ is the time length (duration) of $VM_j$ running on $PM_i$.\\
(2). The total energy consumption of a Cloud Data center (CDC) is computed as\\
\begin{equation}
E_{CDC}=\sum_{i=1}^{n} E_i %~~~~~~~~~~~~~~~~~~~~~~~~~~~~~~~~~~~~~~~~~~~~~~~(13)\\
\end{equation}
It is the sum of energy consumed by all PMs in a CDC. Note that the energy consumption of all VMs on all PMs is included. The objective of our research is to minimize total energy consumption by considering time and capacity constraints. Based on previous results, we have the following results.\\
%This can be proved by an example shown in Figure 2. Optimal solution uses $g+1$ machines and MFFDE algorithm uses $g$ machines, %but the total cost of OPT is less than the total cost of MFFDE.\\

\textbf{Theorem 1. With the allocation function suggested in Equation (1), VMs are allocated to the PMs with high CPU utilization within given threshold.}\\

\textbf{Proof:} From Equation (1), we know that the allocation function $f_a(x)$ is a monotonically increasing function with regard to to CPU utilization $x$. Therefore, PMs with higher CPU utilization will have higher probabilities to accept VMs when other conditions are satisfied. This completes the proof. \\

Theorem 1 also indicates that this allocation function can help reduce energy consumption with smaller number of total PMs used. We will validate this later.\\
\textbf{Theorem 2. With the migration function suggested in Equation (2)-(3), VMs are migrated smoothly (asynchronously) within the given lower and upper thresholds.}\\

\textbf{Proof:} From Equation (2)-(3), we know that the migration function $f_m(x)$ is a monotonically decreasing function when CPU utilization is below $T_l$ and is a monotonically increasing function when CPU utilization beyond $T_h$. Therefore VMs in a PM will be migrated smoothly (asynchronously) following the smooth and asynchronous migration function in two different cases controlled by the migration function. This completes the proof. \\

\section{Performance Evaluation}
\label{performance}

\subsection{Algorithms Compared}
We considered three algorithms in this paper:\\
\begin{itemize}

\item
EcoCloud: this algorithm is introduced in \cite{EcoCloud2013}. It has assignment function and migration function respectively. Its allocation function is
\begin{equation}
f(x,p,T)=\frac{x^{p}(T-x)}{M_p}, x\in [0,T] ,
M_p=\frac{p^{p}(T^{p+1})}{(p+1)^{p+1}}
\end{equation}
where $p$ is a shape parameter and migration functions include the lower part 
\begin{equation}
f_{migrate}^{l}=(1-x/T_l)^{\alpha}
\end{equation}
and the upper part
\begin{equation}
f_{migrate}^{h}=(1+(x-1)/(1-T_h))^{\beta}
\end{equation}
The assignment procedure is performed when a client asks the data center to execute a new application. The migration procedure is defined as follows: each server monitors its CPU and RAM utilization using the libraries provided by the virtualization infrastructure (e.g., VMWare or Hyper-V) and checks if it is between two specified thresholds, the lower threshold $T_l$ and the upper threshold $T_h$. When this condition is violated, the server evaluates the corresponding migration probability function to decide the migration. The two kinds of migrations are also referred to as 'low migrations' and 'high migrations'. The shape of the functions can be modulated by tuning the parameters $\alpha$ and $\beta$, which can therefore be used to foster or hinder migrations. The same function is applied to CPU and RAM, but the parameters, $T_l$, $T_h$, $\alpha$ and $\beta$ can have different values for the two resources.
\item
DRS: this algorithm is introduced in \cite{VMWare}. The DRS (with power efficiency called DPM) is already implemented in VMWare VCenter so we directly use it in real test. The basic ideas in DRS (DPM) for allocation is to try to use small number of PMs and keep load balancing among all PMs. As for migration, it predefines the upper bound of utilization. Once overloaded (beyond the upper bound) happens, it will trigger the migration process.
\item
SAVE: this algorithm is our proposed method introduced in previous sections.
\end{itemize}

\subsection{Simulation}

For all the simulation, we consider that PMs have 400GHz CPU, 400G RAM, 800G Disk, 800Mbps bandwidth; and VMs have 1GHz CPU, 1G RAM, 2G Disk and 2Mbps bandwidth. Each PM has $P_{min}$=110 watts, and $P_{max}$=205 watts.
CloudSim \cite{Calheiros2011} is used for performance evaluation. Three test sets are applied. Each test lasts for 6 hours. Each VM has duration range varying from 1 hour to 6 hours. The results are provided in TABLE 1, 2 ,3 respectively.
In TABLE 1, we set the number of PMs as $P$=100 and the number of VMs as $V$=100. Setting DRS as the baseline, EcoCloud and SAVE has total energy consumption 89.20\% and 68.56\% of the baseline, respectively. The total number of migrations are 216, 127 and 176 for DRS, EcoCloud and SAVE though.\\

In TABLE 2, we set the number of PMs as $P$=100 and the number of VMs as $V$=150. Setting DRS as the baseline, EcoCloud and SAVE has total energy consumption 89.32\% and 71.02\% of the baseline, respectively. The total number of migrations are 244, 120 and 242 for DRS, EcoCloud and SAVE though.\\

In TABLE 3, we set the number of PMs as $P$=200 and the number of VMs as $V$=100. Setting DRS as the baseline, EcoCloud and SAVE has total energy consumption 93.86\% and 72.74\% of the baseline, respectively. The total number of migrations are 298, 240 and 312 for DRS, EcoCloud and SAVE though.\\

For all the tested cases, SAVE has better performance with regard to to total energy consumption while it takes more number of migrations than EcoCloud.

\begin{table}
\caption{ Total energy consumption comparison when PMs=100,VMs=100}
\begin{center}
\begin{tabular}{|l|l|l|l|}
\hline item & DRS&EcoCloud& SAVE
\\\hline
\hline Energy (KWh) & 50.72 & 45.24& 34.77 \\
\hline Migrations &216 & 127&176 \\
\hline Relative cost& 100\% &89.20\%&68.56\%\\
\hline
\end{tabular} \\
\end{center}
\end{table}

\begin{table}
\caption{ Total energy consumption comparison when PMs=100,VMs=150}
\begin{center}
\begin{tabular}{|l|l|l|l|}
\hline item & DRS&EcoCloud& SAVE
\\\hline
\hline Energy (KWh) & 74.05 & 66.14& 52.59 \\
\hline Migrations &244 & 120&242 \\
\hline Relative Cost& 100\%&89.32\%&71.02\%\\
\hline
\end{tabular} \\
\end{center}
\end{table}

\begin{table}
\caption{ Total energy consumption comparison when PMs=200,VMs=100}
\begin{center}
\begin{tabular}{|l|l|l|l|}
\hline item & DRS&EcoCloud& SAVE
\\\hline
\hline Energy (KWh) & 48.21 & 45.25& 35.07 \\
\hline Migrations &298 & 240&312 \\
\hline Relative saving&100\%&93.86\%&72.74\%\\
\hline
\end{tabular} \\
\end{center}
\end{table}

\subsection {Tests in real environment}
For real tests, we just implement SAVE in real VMWare environment so that we can compare against it directly. However, we do not implement EcoCloud in VMWare yet.
\begin{figure} [htp!]
\begin{center}
%\hfil
{\includegraphics [width=0.5\textwidth,angle=-0] {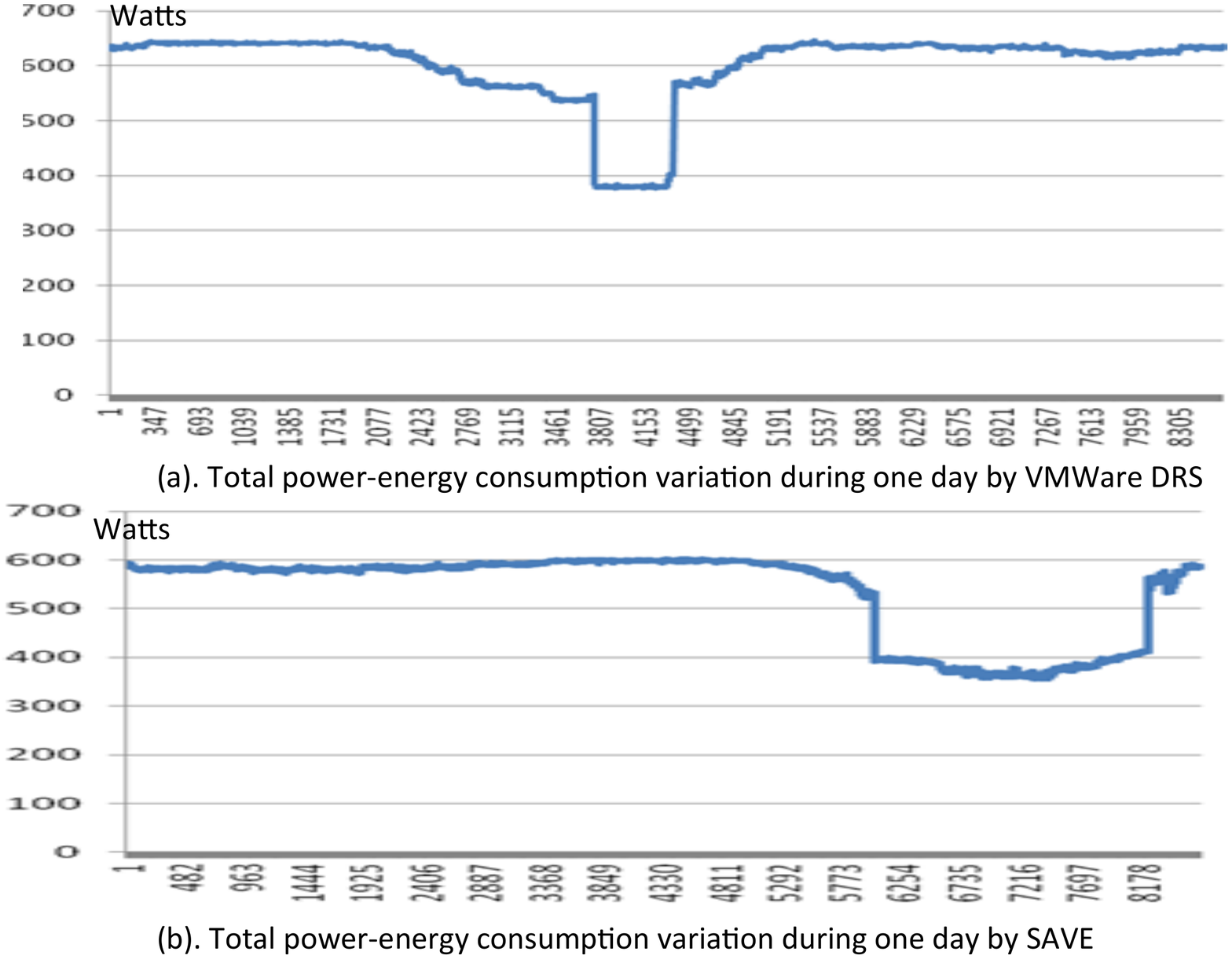}}
%\hspace*{\fill}
\caption{Total Power Comparison between DRS and SAVE}
\end{center}
\end{figure}
We conduct 24 hours real tests respectively for DRS and SAVE, and collect CPU utilization, power consumption, and energy consumption data. In the test, there are three PMs each with 48 CPUs, 40G memory and 1TB disk. Each VM has 1 VCPU, 2G memory and 10GB disk. We generate the same VM requests (22 VMs for each test) for both DRS and SAVE. Each PM has $P_{max}$=210 watts and $P_{min}$=102 watts.\\
1). Firstly we compare the total power consumption between DRS and SAVE. Fig.4 (a) and (b) provide the total power consumption of DRS and SAVE, respectively, where $Y$-axis is for total power consumption (in Watts) and $X$-axis is for the total energy consumption (in watts-hours, WHs). It can be seen that DRS has larger total energy consumption (8305 WHs) while SAVE has total energy consumption of about 8200 WHs during one day test. The reason that DRS has larger total energy consumption lies that it has total power consumption close to 600 Watts during most of time except for interval [11:00, 12:00] while SAVE used only two PMs and saved power and energy consumption during longer intervals of [16:00, 21:00]. \\
\begin{figure} [htp!]
\begin{center}
%\hfil
{\includegraphics [width=0.5\textwidth, angle=-0] {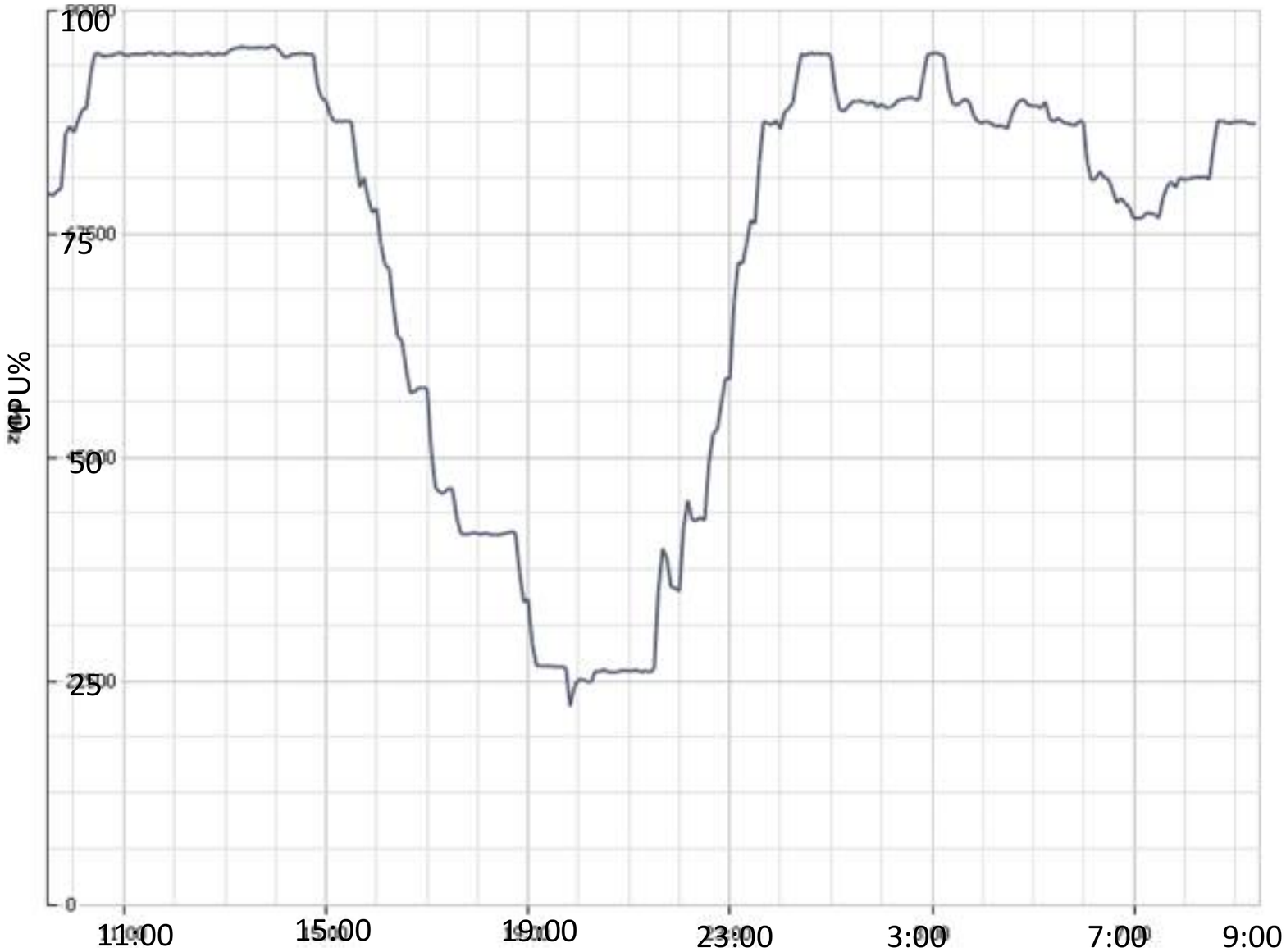}}
%\hspace*{\fill}
\caption{CPU Utilization Variation of DRS in One Day }
\end{center}
\end{figure}
2). Then we compare CPU utilizations between DRS and SAVE. Fig.5 provides average CPU utilization variation of three PMs of DRS in one day, and Fig.6 provides CPU utilization variation of 22 VMs in SAVE in one day. In Fig.5, $Y$-axis is for normalized average CPU utilization which we can collect from VMWare software VCenter directly and $X$-axis is for different time in one day. In Fig. 6, the average CPU of all 22 VMs by SAVE during test are showed. One can see that DRS has lower average CPU utilization, especially during interval [17:00, 23:00] than SAVE which has total CPU utilization between 60-70$\%$ on average. This explains why SAVE is more enerfy-efficient than EcoCloud.  \\
\begin{figure} [htp!]
\begin{center}
%\hfil
{\includegraphics [width=0.5\textwidth,angle=-0] {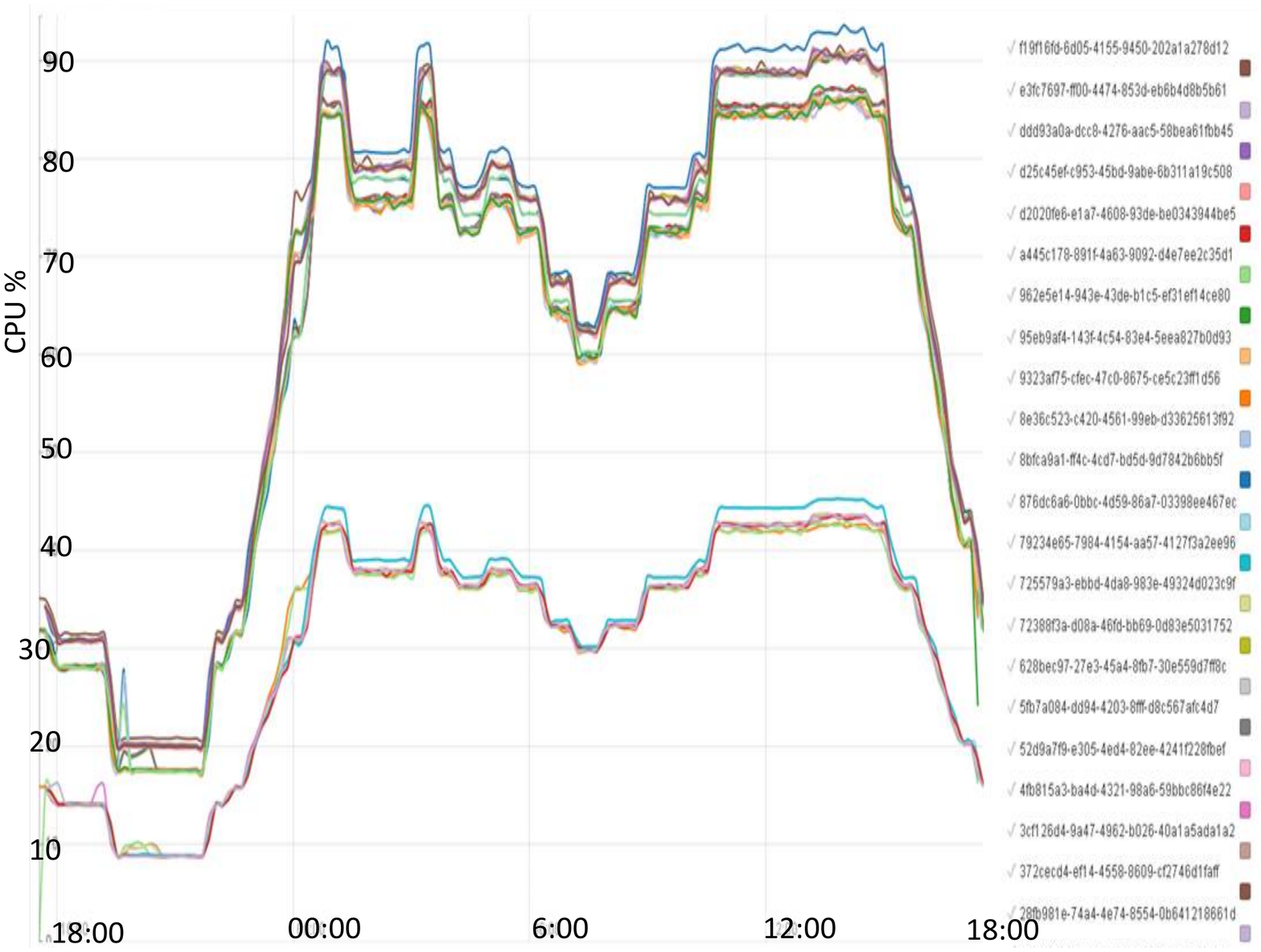}}
%\hspace*{\fill}
\caption{CPU Utilization Variation of SAVE in One Day }
\end{center}
\end{figure}

\section{Related Work}
\label{related work}
For the background and general introduction of cloud computing and energy-efficient scheduling, Beloglazov et al. \cite{Beloglazov2011} propose a taxonomy and survey of energy-efficient data centers and Cloud computing, especially the models for power consumption and energy consumption can be applied. 
%Rimal et al. [?] discuss architectural requirements for Cloud computing systems from an enterprise approach.
Srikantaiah et al. \cite{Srikantaiah2008} study the inter-relationships between energy consumption, resource utilization, and performance of consolidated workloads. Lee et al. \cite{Lee2012} introduce two online heuristic algorithms for energy-efficient utilization of resources in Cloud computing systems by consolidating active tasks. Feller et al. \cite{Feller2012} proposed a novel fully decentralized dynamic VM consolidation schema based on an unstructured peer-to-peer (P2P) network of PMs. Guazzone et al. \cite{Guazzone2011} consider a two-level control model to automatically allocate resources to reduce the energy consumption of web-service applications. \\

For online energy-efficient scheduling, Kim et al. \cite{Kim2011} model a real-time service as a real-time VM request, and use dynamic voltage frequency scaling schemes for provisioning VMs in Cloud Data centers. Tian et al. \cite{Tian2013} propose an online scheduling algorithm for the problem of immediate (on-spot) requests. \\
As for offline energy-efficient scheduling, Beloglazov et al. \cite{Beloglazov2012} consider the off-line VM allocation based on modified best-fit bin packing heuristics without considering VM life cycles, the problem formulation is different from our proposed one. VMware DRS (DPM) \cite{VMWare} and EcoCloud \cite{EcoCloud2013} are two researches closely related to our work, both of them apply distributed solutions. We combine the advantages of centralized and distributed solutions to further we reduce the total energy consumption in a cloud data center.\\

\section{Conclusion and Future work}
\label{conclusion}
In this paper, a self-adaptive energy-efficient scheduling method for virtual machine consolidation is proposed. The approach makes decisions of the assignment and migration of VMs by probabilistic processes and are based on local information, therefore is very simple to implement. Both simulation and real environment tests show that our proposed method SAVE can reduce energy consumption about 30\% against VMWare DRS and 10-20$\%$ against EcoCloud on average. There are a few more open research issues for the problem:
\begin{itemize}
\item
Collecting and analyzing energy consumption data in the real Cloud Data centers. There is still lack of data for open access from real Cloud data centers regarding energy consumption of computing resource. We are evaluating our algorithms in a medium size Cloud Data center to further improve the design.
\item
Combing energy-efficiency and load-balancing together. Just considering energy-efficiency may not be enough for real application because it may cause problems such as unbalance load for each PM. So we will combine load-balancing and energy efficiency together to provide an integrated solution.
\\
We are conducting research to further improve energy efficiency by considering these issues.
\end{itemize}


\section*{Acknowledgment}
This research is sponsored by the National Natural Science Foundation of China (NSFC) (Grand Number:61450110440).



%\bibliographystyle{IEEEtran}

\end{document}